\documentstyle[12pt]{article}

\def\le{\langle}
\def\re{\rangle}

\def\b{\begin{equation}}
\def\e{\end{equation}}

\title{Maximally entangled states and  Bell's inequality in relativistic regime}
\author{Shahpoor Moradi$^1$,
 \thanks{e-mail: shahpoor.moradi@gmail.com} }
\date{\today}

\begin{document}
\maketitle {\it \centerline{$^1$  Department of Physics, Razi
University, Kermanshah, IRAN}}

\begin{abstract}

In this Letter we show that in relativistic regime  maximally
entangled state of two spin-$\frac{1}{2}$ particles not only gives
maximal violation of the Bell-CHSH inequality but also gives the
largest violation attainable for any pairs of four spin
observables that are noncommuting for both systems. Also we extend
our results to three spin-$\frac{1}{2}$ particles. We obtain the
largest eigenvalue of Bell operator and show that this value is
equal to expectation value of Bell operator on GHZ state.

\end{abstract}
\section{Introduction}
Relativistic effects on quantum nonlocality is investigated by
many authors [1-6]. M. Czachor \cite{cz}, investigated
Einstein-Podolsky-Rosen experiment with relativistic massive
spin-$\frac{1}{2}$ particles. The degree of violation of the
Bell's inequality is shown to depend on the velocity of the pair
of  particles with respect to the laboratory.
 He considered the spin singlet of two spin-$\frac{1}{2}$ massive particles moving in the
same direction. He introduced the concept of a relativistic spin
observable using the relativistic center-of-mass operator. For two
observers in the lab frame measuring the spin component of each
particle in the same direction, the expectation value of the joint
spin measurement, i.e., the expectation value of the tensor
product of the relativistic spin observable of each constituent
particle, depends on the boost velocity.

Kar \cite{kar} has shown that a maximally entangled state of two
spin-$\frac{1}{2}$ particles  gives a maximum violation of the
Bell-CHSH inequality. To prove this, Kar made use of a technique
based on the determination of the eigenvectors and eigenvalues of
the associated Bell operator. In this Letter we would like to
extend these results to the relativistic case.

 The paper is organized as
follows: In section 2 we obtain the eigenvalue of Bell operator
for two qubit system. After that we calculate the expectation
value of Bell operator on a maximally entangled state. In section
3 we do the same for three particles case. Finally we conclude
with a discussion in section 4.
\section{Two qubit system}
 For two spin $\frac{1}{2}$-particles, the most commonly discussed Bell's inequality is the CHSH inequality
\b -2\leq\langle{\mathcal{B}}\rangle\leq2, \e where
$\langle{\mathcal{B}}\rangle$ denotes the expectation value of the
Bell-CHSH operator\b
{\mathcal{B}}=\vec{a}.\vec{\sigma}\otimes(\vec{b}+\vec{b}')+\vec{a}'.\vec{\sigma}\otimes(\vec{b}-\vec{b}').
\e Here  $\vec{a}$, $\vec{a}'$, $\vec{b}$ and $\vec{b}'$ are real
three-dimensional vectors of unit length and
$\vec{\sigma}=(\sigma_x,\sigma_y,\sigma_z)$ is the Pauli spin
operator. For each measurement, one of two possible alternative
measurement is performed: $\vec{a}$ or $\vec{a}'$ for particle 1,
$\vec{b}$ or $\vec{b}'$ for particle 2. The square of Bell
operator is given by \cite{la} \b
{\mathcal{B}}^2=4\hat{I}\otimes\hat{I}-[\vec{a}.\vec{\sigma},\vec{a}'.\vec{\sigma}]
\otimes[\vec{b}.\vec{\sigma},\vec{b}'.\vec{\sigma}]. \e With the
help of the identities\b
[\vec{a}.\vec{\sigma},\vec{a}'.\vec{\sigma}]
=2i\vec{\sigma}.(\vec{a}\times\vec{a}')=2i\vec{\sigma}.\vec{c} ,\e
\b [\vec{b}.\vec{\sigma},\vec{b'}.\vec{\sigma}]
=2i\vec{\sigma}.(\vec{b}\times\vec{b}')=2i\vec{\sigma}.\vec{d} ,\e
relation (3) reduces to \cite{kar} \b
{\mathcal{B}}^2=4\left[\hat{I}\otimes\hat{I}+{\sin\theta_{aa'}\sin\theta_{bb'}
{\sigma_{c}}\otimes{\sigma_{d}}}\right],\e where $\theta_{aa'}$ is
the angle between the vectors $\vec{a}$ and $\vec{a}'$,
$\sigma_{c}$ is the spin observable corresponding to a spin
measurement along the unit vector $\vec{c}$ and so on. A
straightforward computation shows that ${\mathcal{B}}^2\leq 8$.
Accordingly, the largest eigenvalue of ${\mathcal{B}}$ is
$2\sqrt{2}$ and the Bell-CHSH inequality can be violated by
quantum state by a maximal factor of $\sqrt{2}$.

 Now we obtain the
relativistic version of (6). The normalized relativistic spin
observable $\hat{a}$ is given by \cite{cz}
\b\hat{a}=\frac{(\sqrt{1-\beta^2}\vec{a}_\bot+\vec{a}_\|).\vec{\sigma}}
{\sqrt{1+\beta^2[(\vec{e}.\vec{a})^2-1]}},\e where the subscripts
$\bot$ and $\|$ denote the components which are perpendicular and
parallel to the boost direction $\vec{\beta}=\beta\vec{e}$.
Operator $\hat{a}$ is related to the Pauli-Lubanski pseudo vector
which is relativistic invariant operator corresponding to spin.
Without loss of generality we assume measurements are in xy-plane
and boost in x-direction. In this case square of Bell operator
takes the form
 \b
{\mathcal{B'}}^2=4\left[\hat{I}\otimes\hat{I}+\frac{(1-\beta^2)\sin(\phi_a-\phi_{a'})\sin(\phi_b-\phi_{b'})
{\sigma_{1z}}\otimes{\sigma_{2z}}}
{\sqrt{(1+\beta^2(a_x^2-1))(1+\beta^2({a'}_x^2-1))(1+\beta^2(b_x^2-1))(1+\beta^2({b'}_x^2-1))}}\right],
\e where we labelled the angles from the x-axis. The eigenstates
are products of eigenstates  of $\sigma_{1z}$ and $\sigma_{2z}$,
which denotes by $|0\re$ and $|1\re$. Here 0 and 1 represent spins
polarized "up" and "down" along the z axis. The largest
eigenvalues of ${\mathcal{B'}}^2$ is given by \b
\zeta=4\left[1+\frac{(1-\beta^2)|\sin(\phi_a-\phi_{a'})\sin(\phi_b-\phi_{b'})|}
{\sqrt{(1+\beta^2(a_x^2-1))(1+\beta^2({a'}_x^2-1))(1+\beta^2(b_x^2-1))(1+\beta^2({b'}_x^2-1))}}\right].
\e The corresponding degenerate eigenstates are $|00\re$ and
$|11\re$ for $\sin(\phi_a-\phi_{a'})$ and $\sin(\phi_b-\phi_{b'})$
having the same sign or $|01\re$ and $|10\re$ for
$\sin(\phi_a-\phi_{a'})$ and $\sin(\phi_b-\phi_{b'})$ of opposite
sign. As every eigenvalue for ${\mathcal{B'}}_H^2$ must lie in the
interval $[0,8]$ it follows that the eigenvalues for
${\mathcal{B'}}$ are necessarily restricted to lie in the interval
$[-2\sqrt{2},2\sqrt{2}]$. It's obvious that in ultrarelativistic
limit as $\beta\rightarrow 1$ Bell's inequality is not violated.
For the following set vector
\[
\vec{a}=\frac{1}{\sqrt{2}}(1,-1),\;\;\;\;\;\;\;\;\vec{b}=(0,1),\]
\b
\vec{a'}=\frac{1}{\sqrt{2}}(-1,-1),\;\;\;\;\;\;\;\;\vec{b'}=(1,0),\e
the square of Bell operator takes the form \b
{\mathcal{B'}}^2=4\left[1+\frac{2\sqrt{1-\beta^2}}{(2-\beta^2)}\right]I\otimes
I, \e then the largest  eigenvalue of ${\mathcal{B'}}_H$ to be
\b\varepsilon_2=\zeta^{1/2}=\frac{2}{\sqrt{2-\beta^2}}(1+\sqrt{1-\beta^2}).\e
In ultrarelativistic limit $\beta\longrightarrow 1$ the amount of
violation is $2$, which indicates Bell's inequality is not
violated. In non relativistic limit $\beta\longrightarrow 0$ we
have the maximum value $2\sqrt{2}$ for $\varepsilon_2$. Here we
obtain identity (12) using the expectation value of Bell operator
on a eigenstate. We assume eigenstate is
\b|\psi\re=\frac{1}{\sqrt{2}}(|00\re+|11\re).\e A straightforward
calculation leads to \b \le \psi|\hat{a}\otimes
\hat{b}|\psi\re=\frac{a_{x}b_{x}+a_zb_z-(1-\beta^2)a_{y}b_{y}}
{\sqrt{(1+\beta^2(a_x^2-1))(1+\beta^2({b}_x^2-1))}}.\e For the set
vector (10) the expectation value of relativistic Bell observable
is exactly (12). The result (12) is obtained by  Ahn, \textit{et
al} \cite{ahn1}. They calculate the Bell observables for entangled
states in the rest frame seen by the observer moving in the $x$
direction and show that the entangled states satisfy the Bell's
inequality when the boost speed approaches the speed of light. The
calculated average of the Bell observable for the Lorentz
transformed entangled states is (12).

\section{Three qubit system}
 Here we consider to three particle case.
For three spin-$\frac{1}{2}$ particles the Bell operator is \b
{\mathcal{B}}_{3}=
\hat{a}\otimes\hat{b'}\otimes\hat{c'}+\hat{a'}\otimes\hat{b}\otimes\hat{c'}+
\hat{a'}\otimes\hat{b'}\otimes\hat{c}-\hat{a}\otimes\hat{b}\otimes\hat{c},\e
where $\hat{a},\hat{a'}$ denote spin observable on the first
qubit, $\hat{b},\hat{b'}$ on the second, and $\hat{c},\hat{c'}$ on
the third. Bell's inequality for three qubits is given by
inequality (1). The square of  Bell operator (13) is given by \b
{\mathcal{B}}_{3}^2=4I-[\hat{a},\hat{a'}][\hat{b},\hat{b'}]-
[\hat{a},\hat{a'}][\hat{c},\hat{c'}]-[\hat{b},\hat{b'}][\hat{c},\hat{c'}]
.\e We assume that three particles move with the same momentums in
x-direction. After some algebra we arrive at \[
{\mathcal{B}}_{3}^2=4\left[\hat{I}\otimes\hat{I}+\frac{(1-\beta^2)\sin(\phi_a-\phi_{a'})\sin(\phi_b-\phi_{b'})
{\sigma_{1z}}\otimes{\sigma_{2z}}}
{\sqrt{(1+\beta^2(a_x^2-1))(1+\beta^2({a'}_x^2-1))(1+\beta^2(b_x^2-1))(1+\beta^2({b'}_x^2-1))}}\right.\]\[\left.
+\frac{(1-\beta^2)\sin(\phi_a-\phi_{a'})\sin(\phi_c-\phi_{c'})
{\sigma_{2z}}\otimes{\sigma_{3z}}}
{\sqrt{(1+\beta^2(a_x^2-1))(1+\beta^2({a'}_x^2-1))(1+\beta^2(c_x^2-1))(1+\beta^2({c'}_x^2-1))}}\right.\]\[\left.
+\frac{(1-\beta^2)\sin(\phi_b-\phi_{b'})\sin(\phi_c-\phi_{c'})
{\sigma_{1z}}\otimes{\sigma_{3z}}}
{\sqrt{(1+\beta^2(b_x^2-1))(1+\beta^2({b'}_x^2-1))(1+\beta^2(c_x^2-1))(1+\beta^2({c'}_x^2-1))}}\right],
\]
we can see that the largest eigenvalue for ${\mathcal{B}}_{3}^2$
is
\[
\lambda_3=4\left[1+\frac{(1-\beta^2)|\sin(\phi_a-\phi_{a'})\sin(\phi_b-\phi_{b'})|}
{\sqrt{(1+\beta^2(a_x^2-1))(1+\beta^2({a'}_x^2-1))(1+\beta^2(b_x^2-1))(1+\beta^2({b'}_x^2-1))}}\right.\]\[\left.
+\frac{(1-\beta^2)|\sin(\phi_a-\phi_{a'})\sin(\phi_c-\phi_{c'})|}
{\sqrt{(1+\beta^2(a_x^2-1))(1+\beta^2({a'}_x^2-1))(1+\beta^2(c_x^2-1))(1+\beta^2({c'}_x^2-1))}}\right.\]\[\left.
+\frac{(1-\beta^2)|\sin(\phi_b-\phi_{b'})\sin(\phi_c-\phi_{c'})|}
{\sqrt{(1+\beta^2(b_x^2-1))(1+\beta^2({b'}_x^2-1))(1+\beta^2(c_x^2-1))(1+\beta^2({c'}_x^2-1))}}\right],
\] which attains maximum value $16$
 with the
following suitably chosen measurement settings,
\[
\hat{a}=\hat{b}=\hat{c}=\hat{y},
\]
\b \hat{a'}=\hat{b'}=\hat{c'}=\hat{x}, \e On the other hand it can
be easily seen that the minimum possible eigenvalue for
${\mathcal{B}}_3^2$ is zero. For example $|001\re$ is an
eigenvector of ${\mathcal{B}}_3^2$ with zero eigenvalue whenever
$\phi_a-\phi_{a'}=\phi_b-\phi_{b'}=\phi_c-\phi_{c'}=\pi/2$. It is
interesting that for three qubits the largest eigenvalue of
${\mathcal{B}}_3$ is not depends on boost velocity which is not
same as two qubit case. Investigations show that exist a family of
pure entangled $N>2$ qubit states that do not violate any Bell's
inequality for N-particle correlations for the case of a standard
Bell experiment on N qubits \cite{gis}. For $N=3$, one class is
Greenberger-Horne-Zeilinger (GHZ) state given by
$|GHZ\re=\frac{1}{\sqrt{2}}(|000\re+|111\re)$. In three qubits
case three experimentalists, Alice, Bob, and Charlotte, can
measure the spin component in arbitrary direction. In
nonrelativistic domain  for a GHZ state, Bell's inequality is
maximally violated if, for example, measurements are made in the
xy-plane along some appropriate directions.  For example with set
vectors (17) and using the algebra of Pauli matrices it is easily
verifiable  for GHZ state Bell's inequality is maximally violated
with value $4$. In relativistic regime again we use the
relativistic spin operator (7) and we assume that three particles
move with the same momentums in x-diraction, then the expectation
values of ${\hat{a}}\otimes {\hat{b}}\otimes{\hat{c}}$ on three
qubit sates are \[ \le 000|{\hat{a}}\otimes
{\hat{b}}\otimes{\hat{c}} |000\re=-\le 111|{\hat{a}}\otimes
{\hat{b}}\otimes{\hat{c}}|111\re=
\]\b
\frac{(1-\beta^2)^{3/2}a_zb_zc_z}{\sqrt{[1+\beta^2(a_x^2-1)][1+\beta^2(b_x^2-1)][1+\beta^2(c_x^2-1)]}}
\e
\[ \le
111|{\hat{a}}\otimes {\hat{b}}\otimes{\hat{c}} |000\re=\le
000|{\hat{a}}\otimes {\hat{b}}\otimes{\hat{c}} |111\re^*=
\]\b
\frac{(a_x+i\sqrt{1-\beta^2}a_y)(b_x+i\sqrt{1-\beta^2}b_y)(c_x+i\sqrt{1-\beta^2}c_y)}
{\sqrt{[1+\beta^2(a_x^2-1)][1+\beta^2(b_x^2-1)][1+\beta^2(c_x^2-1)]}}
\e Then the expectation value on GHZ state is
\[ \le GHZ|{\hat{a}}\otimes {\hat{b}}\otimes{\hat{c}}|GHZ\re\]\b=
\frac{a_xb_xc_x-(1-\beta^2)(a_yb_xc_y+a_yb_yc_x+a_xb_yc_y)}
{\sqrt{[1+\beta^2(a_x^2-1)][1+\beta^2(b_x^2-1)][1+\beta^2(c_x^2-1)]}}\e
So the expectation value of  Bell observable (15) on GHZ state is
4, then Bell's inequality is maximally violated like
nonrelativistic case.

Here we assume  particles are emitted in a plane in a
configuration in which the three momenta lie at angles of $2\pi/3$
to each other. In this situation particles are in the center of
mass frame with the following unite vector boosts  \b
{\mathbf{e}}_1=-\hat{x} ,\e\b
{\mathbf{e}}_2=\frac{1}{2}\hat{x}+\frac{\sqrt{3}}{2}\hat{y}
 ,\e\b
{\mathbf{e}}_3=\frac{1}{2}\hat{x}-\frac{\sqrt{3}}{2}\hat{y},\e
Then the largest eigenvalue of  ${\mathcal{B}}_{3}^2$ to be \b
\lambda=4\left(1+\frac{8\sqrt{1-\beta^2}}{\sqrt{(4-\beta^2)(4-3\beta^2)}}+
\frac{16(1-\beta^2)}{(4-\beta^2)(4-3\beta^2)}\right)
 .\e
therefor we have \b
\varepsilon_3=\lambda^{1/2}=2\left(1+\frac{4\sqrt{1-\beta^2}}{\sqrt{(4-\beta^2)(4-3\beta^2)}}\right)
 .\e
In ultrarelativistic limit as $\beta\rightarrow 1$ approaches to
$2$ and in non relativistic limit $\beta\rightarrow 0$ to be $4$.
which is the maximum value of Bell operator ${\mathcal{B}}_{3}$.
Then similar to two qubit case only in ultrarelativistic limit the
Bell's inequality is satisfied. For set vectors (17) the
relativistic spin operators take the forms
 \b \hat{a}=\hat{y},\;\;\;\;\;\;\hat{a}'=\hat{x} ,\e \b
\hat{b}=\frac{(3+\sqrt{1-\beta^2})\hat{y}+\sqrt{3}(1-\sqrt{1-\beta^2})\hat{x}}{2\sqrt{4-\beta^2}}
,\e
 \b
\hat{c}=\frac{(3+\sqrt{1-\beta^2})\hat{y}-\sqrt{3}(1-\sqrt{1-\beta^2})\hat{x}}{2\sqrt{4-\beta^2}}
,\e
 \b
\hat{b}'=\frac{\sqrt{3}(1-\sqrt{1-\beta^2})\hat{y}+(3+\sqrt{1-\beta^2})\hat{x}}{2\sqrt{4-3\beta^2}}
,\e \b
\hat{c}'=\frac{-\sqrt{3}(1-\sqrt{1-\beta^2})\hat{y}+(3+\sqrt{1-\beta^2})\hat{x}}{2\sqrt{4-3\beta^2}}
,\e One can easily show that for the above spin operators the
expectation value of Bell operator on GHZ state is same as (25).

\section{Conclusions}
We show that the maximally entangled state gives the largest
possible violation for Bell-CHSH  inequality in relativistic
formalism also largest eigenvalue of relativistic Bell operator.
Furthermore, we have shown that the maximal violation of Bell's
inequality predicted by quantum mechanics decreases in
relativistic case. Bell's inequality  in relativistic case is not
always violated, because the degree of violation of Bell's
inequality depends on the velocity of the particles. In non
relativistic case the spin degrees of freedom and momentum degrees
of freedom are independent. But in relativistic regime Lorentz
transformation of spin of particle depends on its momentum.

There are some differences between two and three qubit systems. In
two particle systems when particles move with the same speed,
using the set vector which in non relativistic case yields the
maximum value of Bell operator, we see the relativistic Bell
operator depends on speed of particles. For three qubit case and
the same conditions like two qubit case the amount of violation is
independent of speed. On the other hand when three particles move
in the center of mass frame amount of violation depends on speed
of particles.

\bibliographystyle{plain}

\end{document}